\documentclass[11pt]{article}

\usepackage{epsfig}
\usepackage{graphicx}
\usepackage[tight]{subfigure}
\usepackage{epstopdf}
\usepackage{amsfonts}
\usepackage{amssymb}
\usepackage{amsthm}
\usepackage{amsmath}
\usepackage{multirow}
\usepackage{array}

\newcommand{\newc}{\newcommand}
\newc{\ra}{\rightarrow}
\newc{\lra}{\leftrightarrow}
\newc{\be}{\begin{equation}}
\newc{\ee}{\end{equation}}
\newc{\bs}{\begin{split}}
\newc{\es}{\end{split}}
\newc{\ba}{\begin{eqnarray}}
\newc{\ea}{\end{eqnarray}}
\newc{\ov}{\overline}
\newc{\pa}{\partial}
\newc{\D}{\Delta}

\newc{\nn}{\nonumber}

\newc{\tref}[1]{Table \ref{#1}}
\newc{\eref}[1]{Equation \eqref{#1}}
\newc{\su}[1]{$SU(#1)$}
\newc{\bm}[1]{\mathbf{#1}}
\newc{\fref}[1]{Figure \ref{#1}}
\begin{document}
	\begin{titlepage}
		
		\vspace*{-15mm}
		\begin{flushright}
		CERN-TH-2017-045\\
		\end{flushright}
		\vspace*{0.7cm}

		\begin{center}
{\Large {\bf  % Some observations
On the Octonionic Self Duality equations  of 3-brane  Instantons }}
			\\[12mm]
				Emmanuel~Floratos $^{a,b,c}$
						\footnote{E-mail: \texttt{mflorato@phys.uoa.gr}}
						 \;and\;
			George~K.~Leontaris$^{c,d}$
			\footnote{E-mail: \texttt{leonta@uoi.gr}}
			\\[-2mm]
			
		\end{center}
		\vspace*{0.2cm}
						\centerline{$^{a}$ \it
					 Institute of Nuclear Physics, NRCS Demokritos,}
						\centerline{\it
							Athens, Greece}
			\vspace*{0.2cm}
				\centerline{$^{b}$ \it
			 Department of Physics, University of Athens,}
				\centerline{\it
					Athens, Greece}
				\vspace*{0.20cm}
		\centerline{$^{c}$ \it
			Theory Department, CERN,}
		\centerline{\it
			CH-1211, Geneva 23, Switzerland }
		\vspace*{0.15cm}
			\centerline{$^{d}$ \it
					Physics Department, Theory Division, Ioannina University,}
				\centerline{\it
					GR-45110 Ioannina, Greece }		
		\begin{abstract}
			\noindent
 We study the octonionic selfduality equations  for $p=3$-branes in the light cone gauge and we construct explicitly, instanton solutions for spherical and toroidal topologies in various flat spacetime dimensions $(D=5+1,7+1,8+1,9+1)$, extending previous results for $p=2$ membranes. Assuming factorization of time  we reduce the self-duality equations to  integrable systems and we determine explicitly periodic, in Euclidean time, solutions  in terms of the elliptic functions.  These solutions describe 4d associative and non-associative calibrations in $D=7,8$ dimensions.  It turns out that for spherical topology the calibration is non compact while for the toroidal topology  is compact. We discuss possible applications of our results to the problem of  3-brane  topology change and its implications for a non-perturbative definition of the  3-brane interactions.
	\end{abstract}
\end{titlepage}

\newpage

 \section{Introduction}

  The tremendous progress of  understanding of perturbative superstring theory as well as its duality symmetries in various spacetime backgrounds~\cite{Polchinski} has led to a well substantiated proposal of M theory, the unifying theory of all superstring
  theories~\cite{Duff:1987bx}-\cite{Taylor:2001vb}.
  The new  objects contained in M-theory which are solitonic gravitational back-reactions of various D-branes, are the M2 and M5-branes.
  These objects were expected naturally from the eleven dimensional (11d) supergravity and, in this framework, they are
   considered as fundamental as the strings are for the various 10d supergravities.  The basic obstacle in understanding these objects as fundamentals, lies in  the absence of a coupling constant and the consequent problem of the definition of their self-interactions.

 An interesting proposal to define the self-interactions of the branes is to use their Euclidean instantons to interpolate between vacua (asymptotic states) with different number of branes~\cite{Gibbons:1993sv}. The simplest ones could be instantons interpolating between states of one and two branes. The study of the simplest such instantons is already a difficult problem but one hopes that finding explicit solutions and trying to understand their moduli space would be and interesting approach. The classification of all such instantons and the geometry of their moduli space is probably beyond the present day capabilities.

 Almost three decades ago the problem of determining the corresponding self-duality equations for membranes has been solved in the case of
 $ 2+1$ dimensional embedding spacetime in ref~\cite{Biran:1987ae} and in the case of $4+1$ in ref~\cite{Floratos:1989hf}. In the latter, using results from the study of Nahm equations~\cite{Nahm:1979yw} for Yang-Mills (YM) monopoles it was shown that the self-duality equations form an integrable model and its Lax  pair and conserved quantities were determined. An important work by Ward~\cite{Ward} clarified further the situation reducing the self-duality  equations to Laplace equation in 3d flat space, for the time function of the surface. This function is a level set function -as it is called  in Morse theory- and the problem of determining topology changing membrane instantons is reduced into the search for non-trivial (multi-valued) time functions.

Recent progress in this direction has been made  in the works~\cite{Kovacs:2015xha,Berenstein:2015pxa}. The classification of  self-duality (SD) equations  in various dimensions  has shown that the possible SD equations are determined by the existence of cross products
of vectors which in turn is related to the four division algebras~\cite{Dundarer:1983fe}. Apart from the trivial cases of  branes of dimension $p$ embedded in $p+1$ space dimensions, the interesting cases concern the $p=2$ brane (membrane) and the $p=3$ brane.

 In references \cite{Floratos:2002ga} and \cite{Floratos:1997bi} the study of membrane instanton equations in higher
 dimensions has provided some insight in the  difficulty of the problem and certain explicit solutions have been obtained
 (see also references~\cite{Sfetsos:2001ku}-\cite{Ivanov2006}).
 In the present work we extend the study to the case of 3-branes in $d=8$ spatial dimensions using the octonionic cross product for
 three 8-dimensional vectors.  We obtain a convenient form of a four complex set for the SD equations which enable us to reduce the
 SD equations in six dimensions.

 The SD equations imply the Euclidean second order equations for the 3-branes and satisfy automatically the Gauss Law of volume preserving diffeomorphisms symmetry of the theory. In the case of external fluxes in the theory such as the one coming from $pp$-wave  supersymmetric gravitational backgrounds, it is possible to redefine the time and to reduce the  SD equations to the case of flat background, and find explicit solutions for the 3-sphere and the 3-torus. In the case of the sphere the instantons are periodic in Euclidean time going from a finite radius to infinity and back in finite time, while in the case of  the three torus the periodic solution interpolates between finite radii.

  The layout of this paper is as follows. In section 2 we review the Hamiltonian formalism and the Equations of Motion (EoM) for 3-branes
  in the light cone gauge and in flat spacetime, as well as,  the corresponding Gauss Law. We discuss the symmetry in this gauge which is
  the volume preserving diffeomorphisms of the 3-brane and which gives the possibility to describe the 3-brane as an incompressible fluid.
  In sections 3 and 4, we derive the first order self-duality equations in eight dimensions and its various lower dimensional reductions and we present them in a very useful and suggestive complex form in four-dimensional complex space. We show that these equation imply  the second order equations and the Gauss Law in Euclidean signature spacetime. In section 5,   assuming factorization of time, we solve analytically the 3-brane self-duality equations for the case of spherical ($S^3$) and toroidal ($T^3$) topology of the brane.
  Finally, in section 6 we present our conclusions and discuss the application of our methods to the issue of topology change of three branes, a problem which is relevant to cosmological brane models.

\section{ $S^3$ and $T^3$ branes in the light-cone gauge}

In this section we briefly present the Hamiltonian system in terms of the Nambu 3-brackets. For the
$S^3$ brane in $9+1$ Minkowski spacetime the Hamiltonian in the light-cone gauge is
\be
{\cal H}= \frac{T_3}{2} \int d\Omega_3 \left(\dot{X}_i^2+\frac{1}{3!}\{X_i, X_j, X_k\}^2\right)
\label{Ham}
\ee
where the indices run the $1,2,\dots ,8$ transverse dimensions.
The corresponding EoM are
\be
\ddot{X}_i = \frac{1}{2}\{\{X_i, X_j, X_k\}, X_j, X_k\}~,
\ee
and the Gauss Law takes the form
\[   \{\dot X_i, X_i\}_{\xi_{\alpha},\xi_{\beta}}=0,\; \alpha, \beta=\theta_{1,2,3},\; \alpha\ne \beta  \]
The volume element is
\be
d\Omega_3 =\sin\theta_2 \sin^2\theta_3 d\theta_1d\theta_2d\theta_3~,
\ee
and the Nabu 3-bracket for $S^3$ is defined as follows
\be
\{X_i, X_j, X_k\} = \frac{1}{\sin\theta_2\sin^2\theta_3 }\epsilon^{\alpha\beta\gamma}
\partial_{\alpha}X_i
\partial_{\beta}X_j
\partial_{\gamma}X_k~\cdot
\ee
For $S^3$ there are four functions, namely the polar coordinates of the unit four-vectors,
\ba
 e_1=\cos\theta_1\sin\theta_2\sin\theta_3\,,\;
 e_2=\sin\theta_1\sin\theta_2\sin\theta_3\,,\;
 e_3=\cos\theta_2\sin\theta_3\,,\;
 e_4=\cos\theta_3~\cdot\label{R4v}
\ea
They satisfy the $SO(4)$ relations
\be
\{e_a,e_b,e_c\} = \epsilon^{abcd}e_d, \label{SO4}
\ee
where the indices take the values $1,2,3,4$.
These equations are instrumental for the factorization of the time from the internal coordinates of
the 3-brane.

It is possible to write down explicitly the infinite dimensional algebra of the volume preserving diffeomorphisms $S^3$,
\ba
\{Y_{\alpha}, Y_{\beta}, Y_{\gamma}  \}&=&f^{\delta}_{\alpha\beta\gamma}Y_{\delta}\,,
\ea
using as a basis the hyper-spherical harmonics
\ba
Y_{\alpha}(\Omega)&=&Y_{nlm}(\theta_1,\theta_2, \theta_3), \; \alpha =(nlm),\; m=-l,\dots, l,\; l=0,1,2,\dots, n-1\,.
\ea
The structure constants $f^{\delta}_{\alpha\beta\gamma}$ can be expressed in terms of the 6-$j$  symbols of $SU(2)$
since $SO(4) \simeq SU(2)\times SU(2)$.

When the brane has the toroidal topology $T^3$, the global symmetries are $U(1)\times U(1)\times U(1)$ and three translations
along the cycles of the torus. The basis of functions on $T^3$ is taken to be
\be
e_{\vec n}=e^{i\vec n\cdot \vec{\xi}},\;\vec{\xi}\in [0, 2\pi]^3, \; \vec n =(n_1,n_2,n_3)\in \mathbb{Z}~\cdot
\ee
This basis   defines an infinite dimensional volume preserving diffeomorphism group of the torus
through the Nabu-Lie 3-algebra
\be
\{ e_{\vec n_1}, e_{\vec n_2}, e_{\vec n_3}\}= -i\, {\rm det} (\vec n_1,\vec n_2,\vec n_3 )\,
e_{\vec n_1+\vec n_2+\vec n_3}~\cdot
\ee
It is important to notice that $\forall A\in SL_3(\mathbb{Z})$ the above algebra remains invariant under the
transformations $\vec n\to A\vec n$. Thus,
for this 3-dimensional brane there is an important discrete symmetry, namely the 3-dimensional modular group $SL_3(\mathbb{Z})$
which could have implications in the quantum mechanical spectrum of this object.
The EoM for the 3-torus are formally the same as in $S^3$ but it is much more convenient to complexify the
eight transverse coordinates to four complex ones as we shall see later in the next section.

\section{ Self-Duality and Octonions }

It is known that  $p$-fold, $n$-dimensional vector cross products are defined only in the following cases
\ba
{\rm any} \; p=1,2,3,\dots && n=p+1\nn\\
p=2&&n=7\nn\\
p=3&&n=7\nn\\
p=3&&n=8\,.\label{crpro}
\ea
In the above cases, for any set of $p$ vectors, $v\in \mathbb{R}^n$, their $p$-fold cross product, which we denote by
$\Pi_p(v_1,v_2,\dots , v_p)$,  is a linear map which satisfies the following properties
\ba
\Pi_p(v_1,v_2,\dots , v_p)\cdot v_k &=&0,\; k=1,2,3,\dots , p\,,
\ea
while the norm of the cross product satisfies the important property,
\ba
\left|\Pi_p(v_1,v_2,\dots , v_p)\right|^2&=& {\rm det}(v_i\cdot v_j)~\cdot\label{NCP}
\ea
In the present work we will focus on the last case $p=3, n=8$ of (\ref{crpro}) since the other cases can be obtained
by appropriate reductions to lower dimensions. Now, we proceed to derive the light-cone self-duality equations
for 3-branes living in eight transverse flat dimensions. We note in passing that it is possible after toroidal compactification
to obtain interesting equations for charged 3-branes in lower dimensions.

It is easy to observe that the potential energy term of  the Hamiltonian~(\ref{Ham})  can be rewritten in terms
of the determinant of the induced metric
\[  {\rm det} \left(\partial_{\alpha}X_i\partial_{\beta}X_i\right) =\frac{1}{3!}\sum_{i,j,k=1}^8 \{X_i,X_j,X_k\}^2\,.\]
The cross product of three 8-dimensional vectors of case (\ref{crpro}) is explicitly given as~\cite{Dundarer:1983fe}
\be
\Pi_i(v_1,v_2, v_3)\, :=\,\phi_{ijkl}v_1^jv_2^kv_3^l, \;\;{i,j,k,l}=1,2,\cdots 8~\cdot \label{NCP3}
\ee
where the definition of $\phi_{ijkl}$ and conventions used in the paper are given in the appendix.
Setting $v_{\alpha}^i=\partial_{\alpha}X_i$ where $\alpha=1,2,3$ and $i=1,2,\dots, 8$, we find that
the potential energy of the 3-brane is the norm squared of the above defined cross product of its
tangent vectors (see eq (\ref{NCP})~).
 Another way to see that is to use directly the identity (\ref{PhiPhi}) given in the appendix  for the tensor $\phi_{ijkl}$.
It is obvious now that the Hamiltonian~(\ref{Ham}) of the 3-brane, can be written as
\ba
{\cal H}= \frac{T_3}{2} \int d\Omega_3 \left(\dot{X}_i^2+\frac{1}{3!}\{X_i, X_j, X_k\}^2\right)
        \equiv   \frac{T_3}{2} \int d\Omega_3 \left(\dot{X}_k-i \Pi_k\right)\left(\dot{X}_k+i \Pi_k\right)\,,
\label{Ham2}
\ea
where $i,j,k=1,2,\dots , 8$ and
\[  \Pi_k \equiv \Pi_k(\partial_1X,\partial_2 X, \partial_3X)\,.\]
For vacuum configurations, ${\cal H}=0$, and in Euclidean time we find the self-duality equations
\ba
\dot{X}_i =\pm \frac{1}{3!}\phi_{ijkl}\{X_j, X_k, X_l\}~\cdot \label{SD8}
\ea

\section{The  3-brane instantons in various Dimensions}

There is a natural  generalization of the self-duality equations in 8 dimensions in terms of the symbol
$\phi_{ijkl}$,
\ba
\dot{X}_i = \frac{1}{3!}\phi_{ijkl}\{X_j, X_k, X_l\}\,. \label{SD88}
\ea
where now the indices $i,j,k,l$ run for 1 to 8. For convenience, we introduce the shorthand notation
\be
X_{ijk}\equiv \{X_i, X_j, X_k\}\label{notijk}
\ee
where the indices run from 1 to 8. We know however that the same equation exists in seven dimensions that is,
$i,j,k,l=1,2,\dots 7$, because of the existence of the cross product $p=3, n=7$ given  in section 3.
Then, employing the definition of $\phi_{ijkl}$ given in the appendix, we find that
eq (\ref{SD88}) implies  the following eight  non-linear (first order) differential equations
\ba
\dot{X}_1&=& X_{823}+X_{865}+X_{847}+X_{735}+X_{762}+X_{524}+X_{346}\nn\\
\dot{X}_2&=& X_{183}+X_{684}+X_{857}+X_{743}+X_{167}+X_{635}+X_{154}\nn\\
\dot{X}_3&=& X_{812}+X_{485}+X_{867}+X_{472}+X_{751}+X_{265}+X_{416}\nn\\
\dot{X}_4&=& X_{862}+X_{835}+X_{871}+X_{567}+X_{732}+X_{512}+X_{136}\label{SDA}\\
\dot{X}_5&=& X_{816}+X_{845}+X_{872}+X_{647}+X_{371}+X_{623}+X_{146}\nn\\
\dot{X}_6&=& X_{851}+X_{824}+X_{873}+X_{457}+X_{172}+X_{325}+X_{314}\nn\\
\dot{X}_7&=& X_{814}+X_{836}+X_{825}+X_{546}+X_{531}+X_{432}+X_{612}\nn\\
\dot{X}_8&=& X_{321}+X_{615}+X_{642}+X_{534}+X_{174}+X_{376}+X_{275}~\cdot\nn
\ea
We notice that the self-duality equations of 3-branes in seven dimensions mentioned before, are simply
obtained from the above system, by eliminating the last equation and the terms containing the index 8
on the right-hand side of the remaining seven equations.

Next, we proceed to the  complexification  of the general system  (\ref{SDA}) by choosing specific
pairing of the 8 real coordinates, as follows
\be
 Z_1= X_1+i X_4,\;  Z_2= X_2+i X_5,\;  Z_3= X_3+i X_6,\;  Z_4= X_8+i X_7~\cdot
\label{X2Z}
\ee
After some elaborate manipulation of the original system of  equations,
we arrive at the following simple  form
\ba
\dot{Z}_1&=&-\frac 12 \{Z_1,Z_k,\bar Z_k\}+ \{\bar Z_2,\bar Z_3,\bar Z_4\}\label{CE1}
\\
\dot{Z}_2&=&-\frac 12 \{Z_2,Z_k,\bar Z_k\}+ \{\bar Z_3,\bar Z_4,\bar Z_1\}\label{CE2}
\\
\dot{Z}_3&=&-\frac 12 \{Z_3,Z_k,\bar Z_k\}+ \{\bar Z_4,\bar Z_1,\bar Z_2\}\label{CE3}
\\
\dot{Z}_4&=&-\frac 12 \{Z_4,Z_k,\bar Z_k\}+ \{\bar Z_1,\bar Z_2,\bar Z_3\}\label{CE4}
\ea
We observe that the  equations (\ref{CE1}-\ref{CE4}) have an $SU(4)$ symmetry acting on the complex vector $(Z_1,Z_2,Z_3,Z_4)$.
It is possible from the four complex equations to obtain various interesting reductions to lower dimensions.
For example, if we demand that $Z_k, k=1,2,3,4 $ are real (i.e., $X_4=X_5=X_6=X_7=0)$, or if $Z_k$ are
all pure imaginary, that $X_1=X_2=X_3=X_8=0$, then we obtain the self-duality equation of three branes
in four real dimensions
\be
\dot{Y}_i=\frac{1}{3!}\epsilon_{ijkl}\left\{Y_j, Y_k, Y_l \right\}~,
\ee
where $Y_k$ represent  the non-zero coordinates of $Z_k$  for each of the above  two cases.

A more detailed study  of   toroidal compactifications and  double dimension reduction
 to lower dimensional branes, as well as their relation to extended continuous
 Toda systems, will be discussed in a forthcoming work~\cite{Floratos:2017}.

\section{Explicit solutions for 3-brane instantons in eight  dimensions}

We now study the solutions of the four complex non-linear differential equations~(\ref{CE1}-\ref{CE4}).
This system can be considerably simplified if we seek solutions with time factorization
\be
 Z_a= \zeta_a(t) f_a(\sigma_1,\sigma_2,\sigma_3)\,.\label{Ansatz}
 \ee
In the subsequent analysis, we work out the cases of spherical and toroidal topology.

\subsection{Spherical Topology}
For  spherical topology of the closed 3-brane, $S^3$, we  choose the functions $f_a$
to be $f(a)=e_a, a=1,2,3,4$, where $e_a$ are the polar coordinates of unit 4-vectors~(\ref{R4v})
in $R^4$.

In this factorization, the first term of the right-hand side of the complex equations~(\ref{CE1}-\ref{CE4})
is proportional to $\{e_i,e_k,e_k\}$
which is identically zero.  The simplified equations now read
\ba
\dot\zeta_1&=&\bar{\zeta}_2\bar{\zeta}_3\bar{\zeta}_4\label{cz1}\\
\dot\zeta_2&=&\bar{\zeta}_3\bar{\zeta}_4\bar{\zeta}_1\label{cz2}\\
\dot\zeta_3&=&\bar{\zeta}_4\bar{\zeta}_1\bar{\zeta}_2\label{cz3}\\
\dot\zeta_4&=&\bar{\zeta}_1\bar{\zeta}_2\bar{\zeta}_3\label{cz4}~\cdot
\ea
We notice that the equations are invariant under  a global $U(1)$ symmetry $\zeta_k\to e^{iq_k} \zeta_k$
with  $\sum_{k=1}^4 q_k=0$.

In the following, we  consider the conjugate equations and form the four combinations
\[Q_k= \frac{i}{2}\left(\bar\zeta_k \dot\zeta_k -\dot{\bar\zeta}_k \zeta_k\right) = \frac{i}{2}\left(
\bar{\zeta}_1\bar{\zeta}_2\bar{\zeta}_3\bar{\zeta}_4-{\zeta}_1{\zeta}_2{\zeta}_3{\zeta}_4\right)~,
\]
for $k=1,2,3,4$. Assuming  polar form
\be  \zeta_k = r_k(t) e^{i\phi_k(t)}~,\label{polarf}
\ee
  we find
\ba
r^2_k\dot{\phi}_k&=&Q~,
\ea
where $Q$ is a conserved quantity
\be
Q=-r_1r_2r_3r_4\sin\phi, \;  \phi=\sum_{i=1}^4\phi_i~\cdot
\ee
Furthermore, subtraction for two different values of $k$ gives three new constants of motion
\be
|\zeta_k|^2-|\zeta_l|^2 =c_{kl}~,
\ee
which are the analogue of Euler's equations for the rigid body.  Substituting (\ref{polarf})  in the sums,
$\bar\zeta_k \dot\zeta_k +\dot{\bar\zeta}_k \zeta_k$  we observe
\be
\frac{dr_k^2}{dt}=2 r_1r_2r_3r_4 \cos\phi~\cdot
\ee
For $r_1^2=s$, using  the conservation laws we write this equation  as follows
\be
\dot s=\sqrt{s(s-a)(s-b)(s-c)-Q^2}~\cdot \label{Wei4}
\ee
%%%%%%%%%%%%%%%%%%%%%%%%%%%%%%%%%%%%%%%%%%%
%   Fig 1
%%%%%%%%%%%%%%%%%%%%%%%%%%%%%%%%%%%%%%%%%%%%
\begin{figure}[t]
	\begin{center}
		\includegraphics[width=0.465\textwidth,angle=0,scale=0.8]{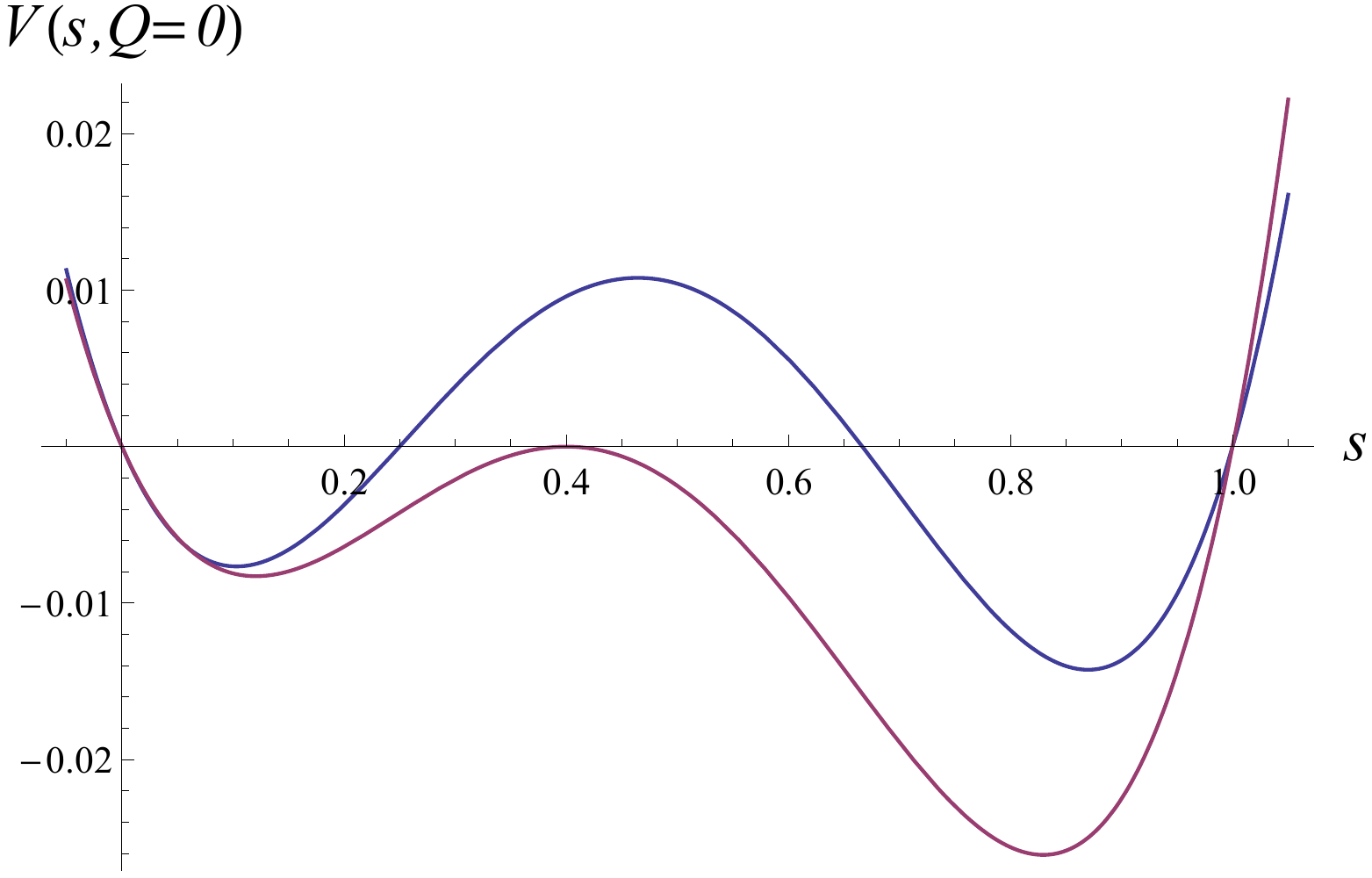}
		\hspace{0.1cm}
		\includegraphics[width=0.465\textwidth,angle=0,scale=.7]{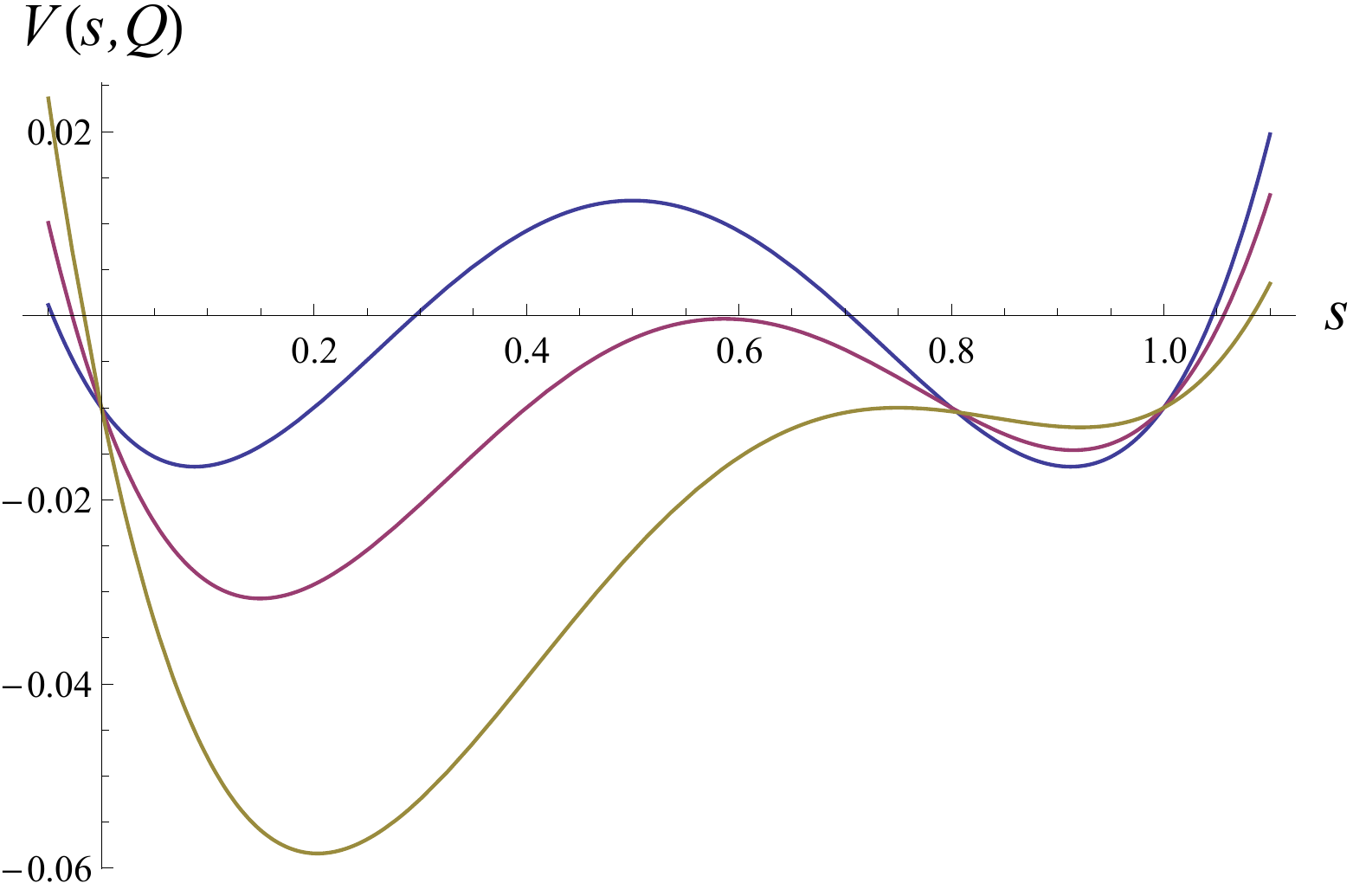}
	\end{center}
	\caption{The `potential' for various values of the radii,  $Q=0$ (left) and $Q=Q_0\ne 0$ (right) }
	\label{fig:1}
\end{figure}
This equation can be solved using standard elliptic functions. Depending on the initial conditions,
we divide the solutions into two classes: those with $Q=0$ where there is no rotation, and those with $Q\ne 0$
 where we have rotating instantons. In figure~\ref{fig:1} we plot the `effective potential'
(i.e. the square of the right-hand side of the differential equation) for $Q=0$ and $Q\ne 0$.
In the first case (left plot) we show two admissible cases. The upper curve stands for four distinct real roots (radii),
and the second curve corresponds to a double root, i.e., when two radii are equal. Curves below this one
do not cross the real axis and correspond to complex radii which is not admissible.
In the second plot, we fix the value of $Q=Q_0$ and draw three characteristic cases.

\begin{figure}[t]
  	\begin{center}
  	\includegraphics[width=0.465\textwidth,angle=0,scale=.8]{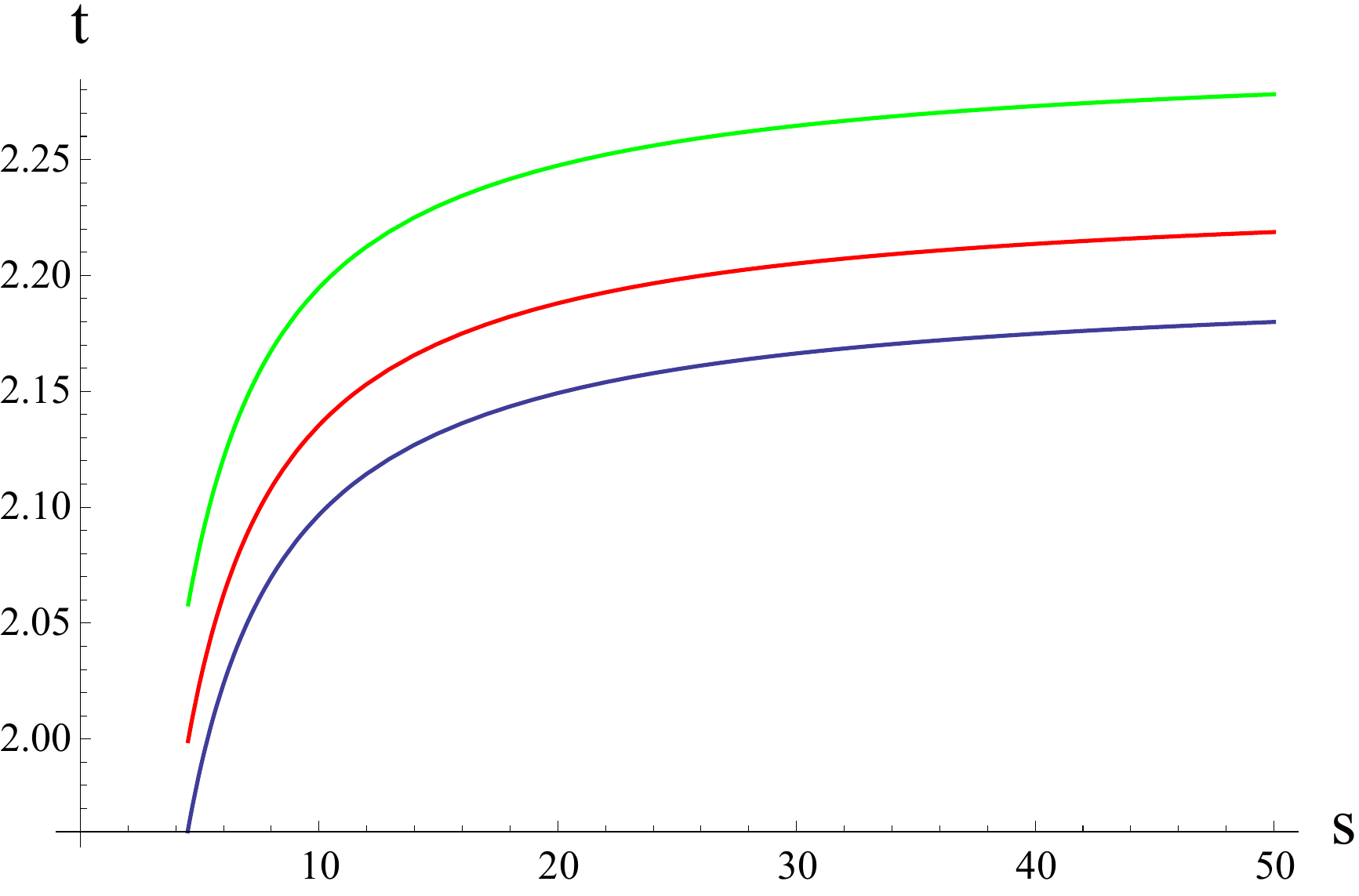}\;\;\;\;\;	
  		\includegraphics[width=0.465\textwidth,angle=0,scale=.8]{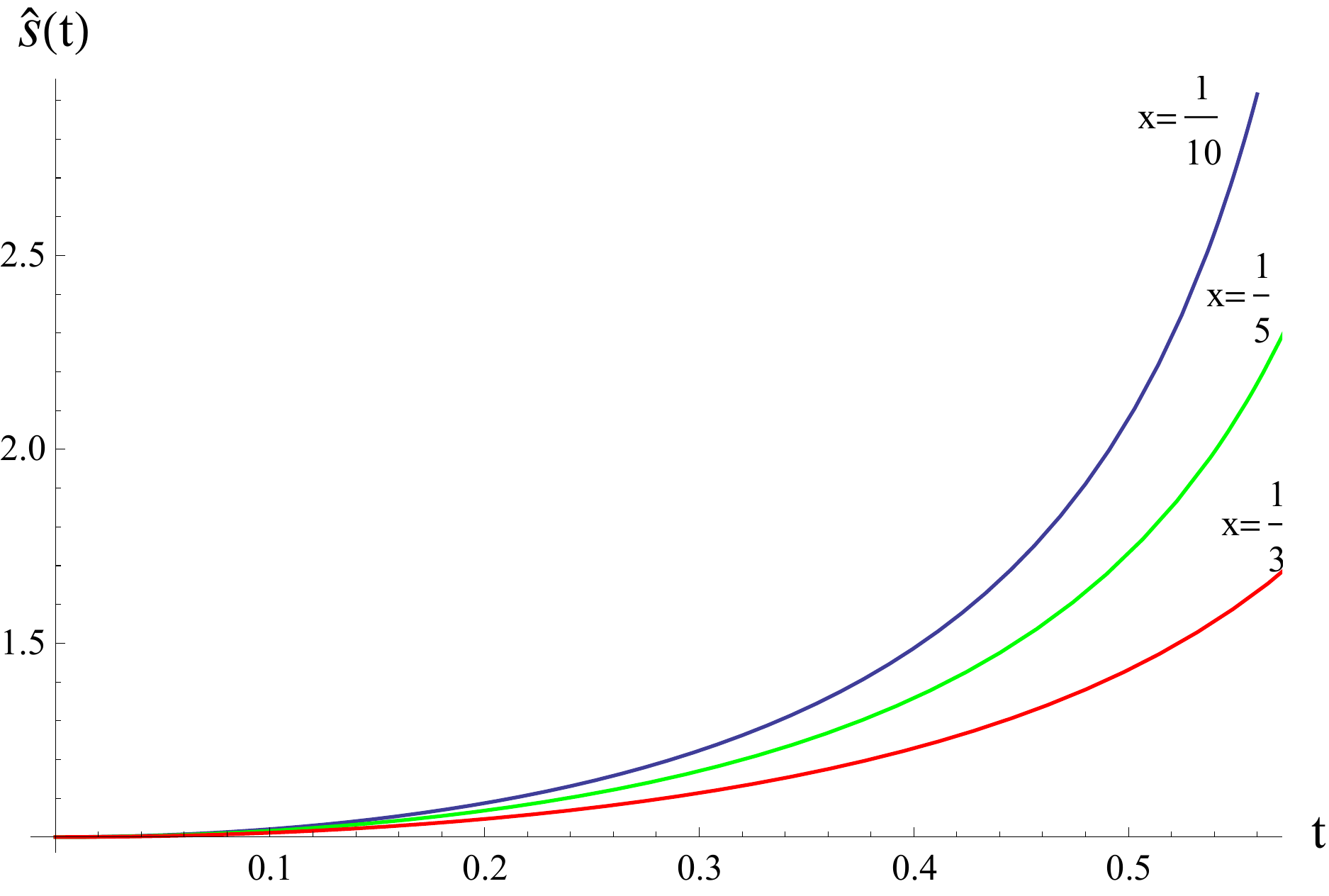}	
  		\end{center}
  	\caption{Left: In the case of spherical topology, the radius goes to infinity at finite time.
  	Right:  plot of $\hat s(t)=\frac{s(t)}{a}$   for $Q=0$ and a double root  $c=b<a$ for
  	the values $x\equiv \frac ba =\frac{1}{10},\frac{1}{5}, \frac 13$ in the case of spherical topology.
  	        }
  	\label{fig:y}
  \end{figure}

First, we consider the case $Q=0$ and the possibility of static solutions, $\dot s=0$. If one of the radii
is zero ($s=a, s=b, s=c$ or $s=0$), then we obtain static solutions  living in the other six dimensions.
In general,  we obtain a solution in terms of the elliptic integral of the first kind $F(\phi|m)$.
In particular, the positivity of all roots requires $s\ge {\rm max}(a,b,c) $.
Assuming $a>b>c>0$ while taking  $s=a$ at the initial time  $t=t_0$,
integration of (\ref{Wei4}) gives
\be
t-t_0 = \frac{2}{\sqrt{c (a-b)}}\left( i K\left(\frac{b (a-c)}{c(a-b)}\right)
- F\left(\sin ^{-1}\left(\sqrt{\frac{(a-b) (s-c)}{(b-c)
   (a-s)}}\right)|\frac{a (c-b)}{c (a-b) }\right)\right)~\cdot \label{timeS}
\ee
In order to find $s(t)$  we  need the inverse of  the elliptic integral of the first kind
  $F(\phi|k)=\int_0^{\sin\phi}\frac{ds}{\sqrt{(1-ks^2)(1-s^2)}}$  which is given by
    the Jacobi Amplitude  $t= F(\phi|k) \to  \phi= F^{-1}(t, k) ={\rm am}(t,k)$.
     Furthermore,  in order to absorb unimportant factors, we redefine time
    ${\tiny t =\frac{2}{\sqrt{c (a-b)}} t' }$  and make the definitions
         \ba
             k&=&\frac{a (c-b)}{c(a-b)}
     \\
        \sigma&=& \sqrt{\frac{(a-b) (c-s)}{(c-b) (a-s)}}=\sin\left( {\rm am}(t,k)\right)~,
        \label{sigt}
        \ea
    (also written in the literature as $ sn(t,k)$)  and finally solve (\ref{sigt}) for $s(t)$.
 In figure~\ref{fig:y} (left side) we plot time as a function of $s=r^2$ for $a=1$ and three sets of
 $b,c$ values. In all cases, the radius goes to infinity at finite time.

 For the sake  of simplicity, we exemplify the above for the case of  a double root  $b=c<a$, i.e,
 when two of the radii are equal.
 Then
   \be
  \int_a^s\frac{du}{\sqrt{u (u-a)} (u-b)}= \frac{2 }{\sqrt{b (a-b)}}
  \left( \frac{\pi}2- \tan^{-1}\left(\sqrt{\frac{s (a-b)}{b (s-a)}}\right) \right)~\cdot
  \ee
  If we define $x=\frac ba <1$, a new time ${\tiny \frac{a t}{\pi}}\to t$, and
  \[q(x,t)= x \tan ^2\left(\frac{1}{2} \pi  \left(1-t \sqrt{(1-x) x}\right)\right)\,,\]
we obtain
  \be
  \hat s(t) \equiv \frac{s(t)}a=\frac{q(x,t)}{x-1+q(x,t)}~\cdot 
  \ee
  This takes its  minimum value for $\hat s(0)=1$ and grows rapidly to infinity in finite time,
  at the zeros of the denominator. A few representative cases are plotted in the right side
  of figure~\ref{fig:y}.
  %%%%%%%%%%%%%%%%%%%%%%%%%%%%%%%%%%%%%%%%%%%
  %   Fig x
  %%%%%%%%%%%%%%%%%%%%%%%%%%%%%%%%%%%%%%%%%%%%
  %%%%%%%%%%%%%%%%%%%%%%%%%%%%%%%%%%%%%%%%%%
The case $Q\ne 0$ can be separated into two classes:  first, pure uniform  rotational motion with
 time independent radii (Euler tops), and second, when we have both rotation and bounce  motion.
 In the second class we consider three different cases according to the initial conditions
 for the radii and the value of $Q$. Without loss of generality we can order the initial values of the
 radii in degreasing magnitude, that is $r_1>r_2>r_3>r_4$, which implies  $a>b>c>0$.
 The solution can be written in terms of the elliptic integral of the first kind, once we use the
 four roots $e_1, e_2, e_3, e_4$ which are  algebraic functions of $a,b,c$ and $Q$. From the geometry of the
 potential plot, we see that when $Q=0$, $e_1=0, e_2=c, e_3=b, e_4=a$. When $Q\ne 0$ but smaller than
 a critical value, $Q_{c}$, the root $e_{1}$ is negative, $e_2$ and $e_4$ increase while $e_3$ decreases.
When $Q=Q_{c}$, two roots become equal,  $e_2=e_3$, and the potential has a local maximum $V(Q_c)=0$.

\subsection{Toroidal Topology}
For the case of toroidal topology, we define
\be
Z_a = \zeta_a(t) e ^{\vec n_a\cdot \vec{\xi}}~\cdot    \label{TAZ}
\ee
The four  vectors $\vec n_a\in \mathbb{Z}^3$ are not completely arbitrary.
When we
substitute (\ref{TAZ}) into the system of the four complex equations~(\ref{cz1}-\ref{cz4}) we find that
the above Ansatz provides a solution only if
\be
\vec n_1+ \vec n_2+ \vec n_3+ \vec n_4 =\,\vec{0}~\cdot
\label{ncond}
\ee
Then we get
\ba
\dot\zeta_1&=&+i\,n
\bar{\zeta}_2\bar{\zeta}_3\bar{\zeta}_4\label{tz1}\\
\dot\zeta_2&=&-i \,n
\bar{\zeta}_3\bar{\zeta}_4\bar{\zeta}_1\label{tz2}\\
\dot\zeta_3&=&+i\,
\bar{\zeta}_4\bar{\zeta}_1\bar{\zeta}_2\label{tz3}\\
\dot\zeta_4&=&-i \,
\bar{\zeta}_1\bar{\zeta}_2\bar{\zeta}_3\,,\label{tz4}
\ea
where $n$ stands for the determinant
\be
n={\rm det }(\vec n_j, \vec n_k, \vec n_l)~.
\ee
From these and their complex conjugates we readily get,
\ba
\frac{d}{dt} |\zeta_1|^2= \frac{d}{dt} |\zeta_3|^2&=&+i n (\bar \zeta_1\bar \zeta_2\bar \zeta_3 \bar \zeta_4-\zeta_1\zeta_2 \zeta_3 \zeta_4)
\\
\frac{d}{dt} |\zeta_2|^2= \frac{d}{dt} |\zeta_4|^2&=&-i n (\bar \zeta_1\bar \zeta_2\bar \zeta_3 \bar \zeta_4-\zeta_1\zeta_2 \zeta_3 \zeta_4)
\cdot
\ea
Subtraction and addition of the appropriate equations gives
\ba
 |\zeta_1|^2-|\zeta_3|^2&=& c_{13}\label{ts1}\\
  |\zeta_2|^2-|\zeta_4|^2&=& c_{24}\label{ts2}\\
  |\zeta_1|^2+|\zeta_2|^2&=& c_{12}\label{ts3}\\
   |\zeta_3|^2+|\zeta_4|^2&=& c_{24}\label{ts4}~\cdot
\ea
We define the currents
\be
Q_a=-\frac{1}{2} \left( \dot{\zeta}_a\bar \zeta_a-\dot{\bar \zeta}_a \zeta_a\right)~,
\ee
and find that
\be
 Q_1=Q_3 =Q\equiv \frac{n}{2} \left(\bar J+J\right),\;  Q_2=Q_4=-Q=-\frac{n}{2} \left(\bar J+J\right)\label{currents}
 \ee
where  $J= \zeta_1\zeta_2 \zeta_3 \zeta_4$ and $\bar J$ its complex conjugate.
We observe now that $\dot{\bar J} = -\dot J$,  which implies
\[ \frac{d Q}{dt}=\frac n2 \frac{d }{dt}(\bar J+J)=0\,,\]
so that the charge $Q$ is conserved.  Writing $\zeta_k = r_k e^{i\phi_k}$,
we find that $Q= r_1r_2r_3r_4\cos\phi$. Furthermore, using the redefinitions $Q'=Q/n$ and $t'= 2 n t$, 
while taking $r_1^2=s$, we obtain the differential equation
\ba
\frac{ds}{dt}&=&\sqrt{s(a-s)(s-b) (c-s)-Q^2}~,
\ea
with $a=c_{12}, b=c_{13}, c=c_{13}+c_{34}$.  This is  exactly the same equation as in the case of spherical topology, but
notice that $r_k$  are bounded form equations (\ref{ts1}-\ref{ts4}).

\noindent
We note that with the same method it is also possible to solve the four-dimensional 3-brane self-duality equations presented before,
where we shall end up with the real system in the place of equations~(\ref{cz1}-\ref{cz4}).
In this case, the system of equations is the elegant integral system of refs~\cite{Corrigan:1982th,Ivanov2006}.

  The solution for the case of toroidal topology is expressed for both cases $Q=0$ and $Q\ne 0$ by the same
elliptic integral as in the spherical topology, but from the conservation laws we find the interesting result
that the motion is oscillatory and bounded and so, the 4-dimensional manifold is compact. This is
 contrary to the case of spherical topology where we have always   non-compact world-volumes.
Since this Euclidean world volumes are calibrations of the ($8+1$)-dimensional embedding spacetime,
our solution is among the first examples of compact non-associative (octonionic) calibrations.

\begin{figure}[t]
	\begin{center}
	\includegraphics[width=0.465\textwidth,angle=0,scale=0.85]{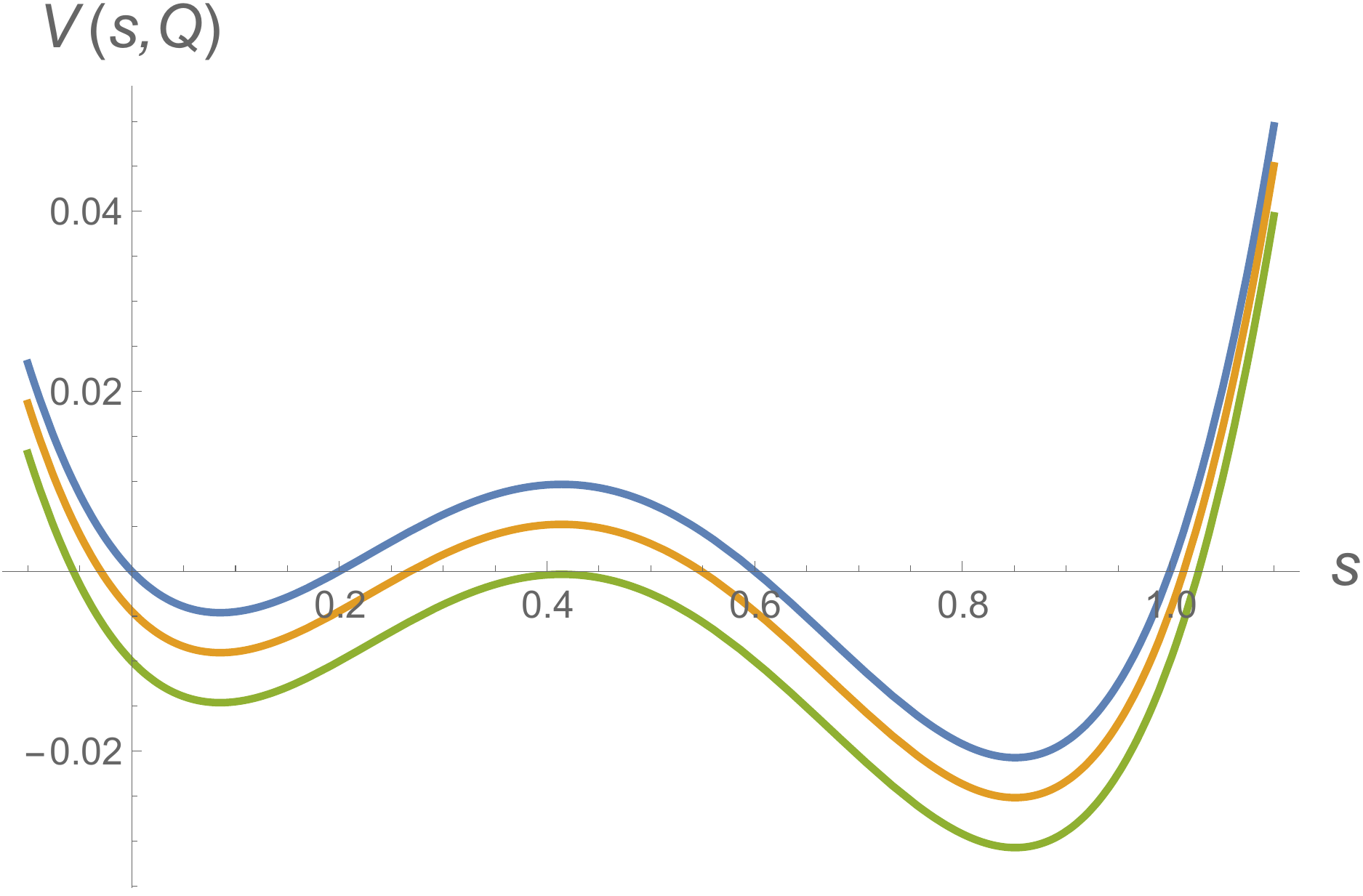}
	\end{center}
	\caption{Time vs radius squared $s=r_1^2$  for  $a>c>b$ and three characteristic $Q$ values,   $Q=0$,
$Q<0$ 	and the critical value $Q=Q_{c}$ for toroidal topology.}
	\label{fig:tt}
\end{figure}

\begin{figure}[t]
	\begin{center}
		\includegraphics[width=0.4\textwidth,angle=0,scale=0.96]{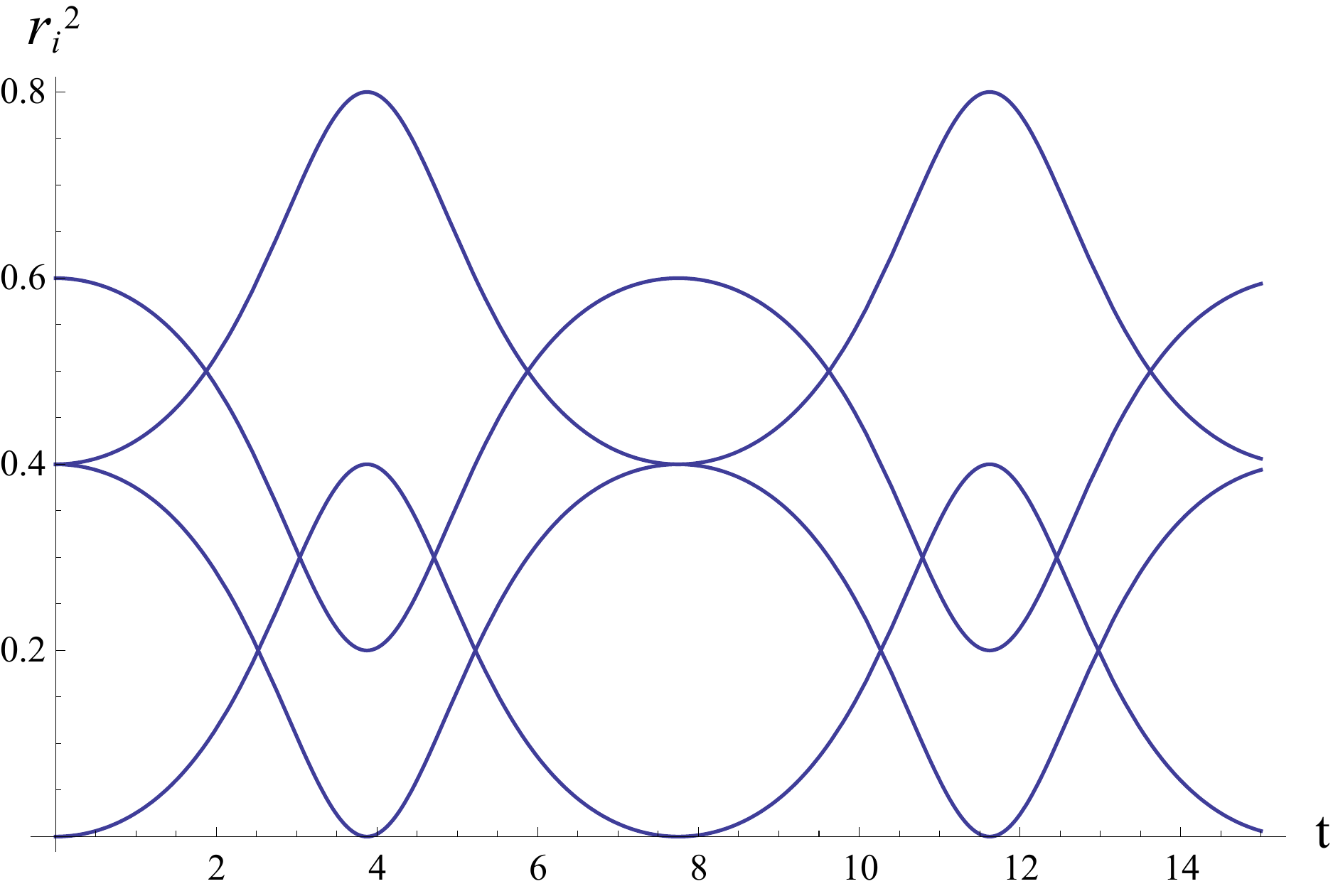}\;\;\;
	%	\includegraphics[width=0.4\textwidth,angle=0,scale=0.58]{radius3.pdf}
	%	\\
%	\includegraphics[width=0.4\textwidth,angle=0,scale=0.6]{radius2.pdf}\;\;\;
	%		\includegraphics[width=0.4\textwidth,angle=0,scale=0.58]{radius4.pdf}
	\end{center}
	\caption{ The plot shows the time evolution of the four radii, for the case $Q=0$
		and the sets $1,1/5,3/5$. Notice that all of the radii are bounded.}
	\label{tor4}
\end{figure}

Indeed, ordering the four radii in decreasing order $r_{k+1}<r_k$,  inspection of the equations (\ref{ts1}-\ref{ts4})
shows that
the constants $a,b,c,d$ must satisfy $b<a<c$ with $a,c$ positive while the variable $s$ is bounded
between $a$ and $b$,  where now $ b<s<a<c$.
In figure~\ref{fig:tt}, the potential is plotted for three characteristic values of $Q$.
The time evolution of the four radii, is also shown in figure~\ref{tor4}. As already emphasized,
all four radii are bounded and interpolate between a minimum and a maximum value.
%%%%%%%%%%%%%%%%%%%%%%%%%%%%%%%%%%%%%%%%%%%%
%   Fig tt
%%%%%%%%%%%%%%%%%%%%%%%%%%%%%%%%%%%%%%%%%%%%
%%%%%%%%%%%%%%%%%%%%%%%%%%%%%%%%%%%%%%%%%%
%   Fig 3
%%%%%%%%%%%%%%%%%%%%%%%%%%%%%%%%%%%%%%%%%%%%
%%%%%%%%%%%%%%%%%%%%%%%%%%%%%%%%%%%%%%%%%

\section{Conclusions and Open Questions}

The present work provides a method for solutions of the self-duality equations for 3-branes in higher dimensions.
The factorization of time exploits the finite sub-algebras of the volume preserving diffeomorphisms  and reduces the SD equations
 to well known integrable systems with explicit solutions in terms of the standard elliptic functions. The new result is that
 the 3-sphere instanton interpolates between flat space-time at infinity and a finite radii 3-sphere. This is similar to
  the Einstein-Rosen wormhole solutions  of General Relativity. The minimization of the Euclidean  4-volume in 8-dimensions,
   which is the origin of the SD equations, classifies their solutions as calibrations of the background geometry.
   For the sphere this calibration is non-compact while for the 3-torus the calibration is compact~\cite{Joyce}.
   It is possible to solve with the same method SD equations in maximally supersymmetric geometry backgrounds such
   as $pp$-waves with fluxes in $8+1$ spacetime dimensions  by redefining the time variable~\cite{Bachas:2000dx},
   a trick which has also a physical implication in the interpretation of the  time as the group renormalization
   scale and the SD equations as flow equations between various geometries~\footnote{We thank Costas Bachas for this observation.}.
     It would be interesting to apply similar ideas with those of ref~\cite{Kovacs:2015xha} to transform the nonlinear
     three-brane SD equations into a linear Laplace equation problem for $8+1$ space time dimensions to
 study possible topology changes~\cite{Floratos:2017}. Another interesting
  direction is to abandon the factorization of time and try to solve the three-brane SD equations in $8+1$ dimensions requiring
  axial symmetry and proceed in analogy with the study of real time BPS states of 3-branes.

\vspace*{.3cm}
{\it Acknowledgments.}\;
The authors   would like to thank CERN theory division for their
kind hospitality and the stimulating atmosphere during which the main part
of this work was realized. EGF would like to thank also the theory
department of ENS in Paris for their kind hospitality and stimulating
atmosphere during which the last parts of this paper were finished.

\newpage
\appendix

\section{The octonionic structure constants}
The structure constants of the octonionic multiplications  $\Psi_{ijk}$ and its dual $ \phi_{ijkl}$,
which measures the non-associativity of octonions, are given by (for a textbook, see for example~\cite{Conway})
\ba
\Psi_{ijk}&=&\left\{\begin{array}{ccccccc}1&2&3&4&5&6&7\\
                             2&4&6&3&7&5&1\\
                             3&6&7&5&2&1&4
 \end{array}\right.\label{Psi}\\
 \phi_{ijkl}&=&
 \left\{\begin{array}{ccccccc}4&3&5&6&1&7&2\\
                              5&7&2&1&3&4&6\\
                              6&5&1&7&4&2&3\\
                              7&1&4&2&6&3&5\\
  \end{array}\right.
\label{Phi}
\ea
Below, we provide the simplest identities between these tensors used in our computations
and by summing more pairs of indices we may obtain additional ones.

\noindent
When two indices are summed, a useful  multiplication rule connecting the two symbols is
\ba
\Psi_{ijk}\Psi_{lmk}=\delta_{il}\delta_{jm}-\delta_{im}\delta_{jl}+\phi_{ijlm}
\label{PsiPhi}
\ea
The corresponding identity for the $\phi_{ijkl}$ symbols is written as follows~\cite{Dundarer:1983fe}
\ba
\phi_{abcl}\phi_{ijkl} &=&\left(\delta_{ai}\delta_{bj}-\delta_{aj}\delta_{bi}\right)\delta_{ck}\nn\\
                      &+& \left(\delta_{bi}\delta_{cj}-\delta_{bj}\delta_{ci}\right)\delta_{ak}\nn\\
                      &+& \left(\delta_{ci}\delta_{aj}-\delta_{cj}\delta_{ai}\right)\delta_{bk}\nn\\
                &+& \phi_{abij}\,\delta_{ck}+ \phi_{bcij}\,\delta_{ak}+ \phi_{caij}\,\delta_{bk}\nn\\
                &+& \phi_{abjk}\,\delta_{ci}+ \phi_{bcjk}\delta_{ai}+ \phi_{cajk}\,\delta_{bi}\nn\\
                &+& \phi_{abki}\,\delta_{cj}+ \phi_{bcki}\,\delta_{aj}+ \phi_{caki}\,\delta_{bj}~\cdot
                \label{PhiPhi}
\ea

\end{document}